\documentclass[11pt]{article} 

\usepackage[utf8]{inputenc} 

\usepackage{geometry} 
\usepackage{graphicx} 

\usepackage{booktabs} 
\usepackage{array} 
\usepackage{paralist} 
\usepackage{verbatim} 
\usepackage{subfig} 
\usepackage{amsmath}
\usepackage{braket}
\usepackage{leftidx} 
\usepackage{appendix}
\usepackage{tikz-feynman}
\usepackage{pgfplots,caption}
\usepackage{authblk} 
\usepackage{hyperref} 

\usepackage{fancyhdr} 
\usepackage{sectsty}
\allsectionsfont{\sffamily\mdseries\upshape} 

\usepackage[nottoc,notlof,notlot]{tocbibind} 
\usepackage[titles,subfigure]{tocloft} 

\newcommand{\angstrom}{\textup{\AA}}

\title{Neutron Production via Electron Capture \\ by Coherent Protons}
\author{ Luca Gamberale}
\affil{LEDA srl, Università Milano Bicocca \\ I-20126 Milano, Italy\\ and \\Quantumatter Inc., USA 
}
\date{July 2022} 

\begin{document}
\maketitle
\begin{abstract}
I consider coherent vibrational states of the quantum plasmas formed by the conduction electrons and protons inside a metal hydride. Such states can interact coherently through weak interaction to produce neutrons at very low energy.
The existence of the vibrational coherent states is supported by a recent theoretical analysis showing that these configurations are characterized by an energy gap of the order of 1 eV per particle compared to the incoherent configurations and are therefore dynamically stable. When excited coherently, these configurations are able to transfer their energy, enabling highly energetic mechanisms. In this paper I show how it is possible to produce neutrons through such a coherent mechanism. The produced neutrons are essentially at rest and remain confined within the metal. The theory developed allows for the theoretical calculation of the production rate of neutrons.
\end{abstract}

\section{Introduction}
\begin{figure}
\centering
\begin{tikzpicture}
\begin{feynman}
\vertex (p1) {\((Z,N)\)}; 
\vertex[below right=2cm and 4.5 cm of p1] (pv); 
\vertex[right=9cm of p1] (p2) {\((Z-1,N+1)\)}; 
\vertex[below right=4cm and 0.5 cm of p1] (e1) {\(e\)}; 
\vertex[right=8cm of e1] (e2) {\(\nu_e\)};
\diagram* { {[edges=fermion]
(p1) -- [very thick] (pv) -- [very thick] (p2), 
(e1) -- [thick] (pv) -- [dotted] (e2)},
};
\end{feynman} 

\end{tikzpicture}
\caption{Feynman diagram for the transmutation of a nucleus with $Z$ protons and $N$ neutrons into a $Z-1$ protons and $N+1$ neutrons plus an electronic neutrino via Electron Capture} 
\label{fig:EC}
\end{figure}
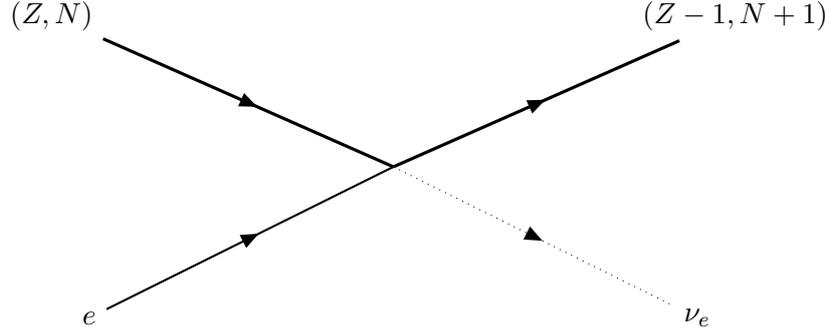

Spontaneous production of neutrons via electron capture (EC)  cannot normally happen in vacuum since the sum of the masses of the proton and the electron is smaller than the mass of the neutron (Fig. \ref{fig:EC}). In order to produce a neutron via electron capture we need an extra-energy of $782$ keV to reach the energy threshold.
The fortunate consequence of this fact is that hydrogen atoms do not spontaneously decay into neutrons, implying that matter is made of atoms and the universe is suitable to support life as we know it. 

However, this well-established result holds true in the perturbative vacuum but may no longer be true in a vacuum with different characteristics.
It is the subject of this paper to show that in fact the EC process
can be influenced by the low energy properties of the surrounding environment.

In the next Sections I analyze the possibility to obtain EC at room temperature and low energy in a metal sample (i.e. Ni, Ti) highly loaded with hydrogen.
The basic idea is that, by means of a coherent process, the energy needed to reach threshold can be supplied by a coherent superposition of energy quanta belonging to highly populated low-energy degrees of freedom, for instance vibrational degrees of freedom of the electron and/or proton plasmas \footnote{It is important to emphasize that in the present paper the vibrational excitations considered are different in nature from those discussed in \cite{metzhag} since in the former case the excitations are the result of a spontaneous symmetry breaking and produce a minimum of the energy of the system whereas in the latter the excitations are supplied by en external forcing drive.}.
For example, assuming that a coherent configuration is composed of $10^{13}$ electrons, a variation of the energy per particle of the order of $8\cdot 10^{-8}$ eV would be sufficient to reach threshold for EC. In the next Sections I will clarify the meaning of the last sentence and analyze how this can be accomplished.

\section{Coherent plasmas}
In \cite{https://doi.org/10.48550/arxiv.2205.14015,ref5} it has been shown that in suitable conditions a collection of electrons or ions at sufficient density condenses spontaneously in coherent configurations to form coherent quantum plasmas where a radiative electromagnetic field is present oscillating in phase with the matter field. Such a quantum state acquires an energy gap and is therefore stable, allowing for coherent scattering processes. This result is valid for bosons as well as for fermions. 

\begin{figure}
\centering
\includegraphics[width=.7\textwidth]{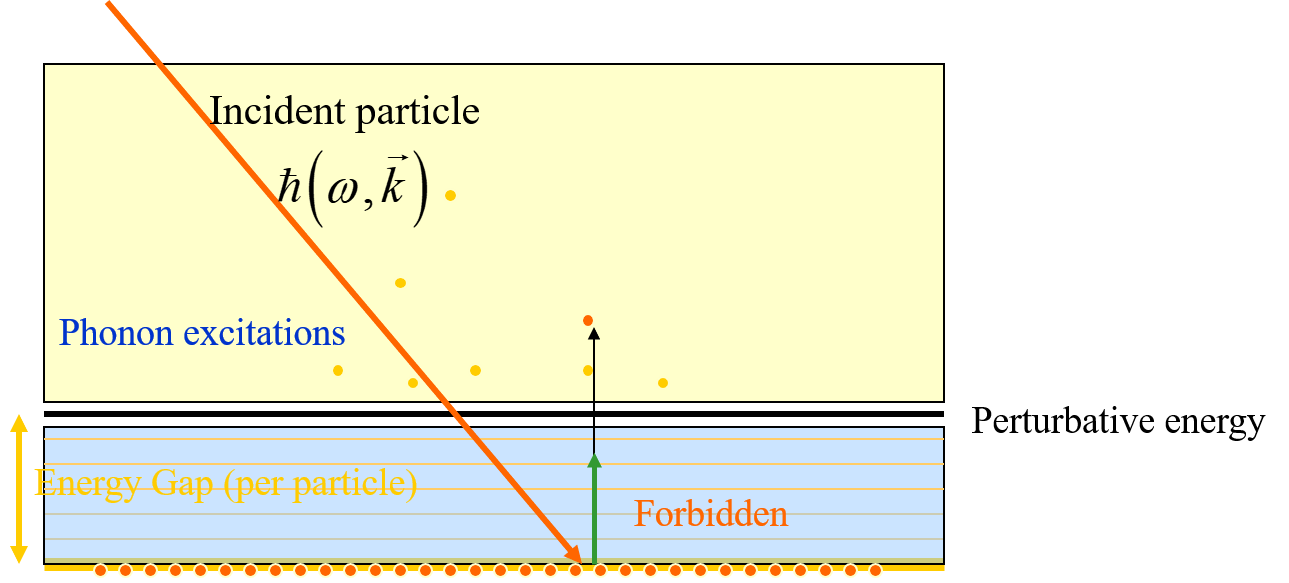}
\caption{Effect of the presence of an energy gap in the kinematics of a scattering process. When the exchanged energy is not sufficient to overcome the energy gap the single-particle interaction cannot happen, leaving room to coherent interactions only.}
\label{fig:coherentscatteringspectrum}
\end{figure}

The key ingredients of the theory are the following:
\begin{itemize}
\item In the metal several quantum plasmas exist, composed of a collection of spatial regions called \emph{Coherence Domains} (CD) whose diameter is 10 $\mu$m on average.
 Within each CD the matter field is in a coherent state, oscillating in 
phase quadrature with the electromagnetic field emitted at unison by the charges composing the quantum state.
\item As a consequence of such an oscillation a spontaneous symmetry breaking occurs so that an {\it order parameter} emerges that keeps in phase all the particles composing the coherent state (Fig. \ref{fig:coherentWF}). The plasma oscillation frequency is shifted ($\omega_R<\omega_p=|\vec k|$, the photon gets a mass) so that a classical EM field remains trapped inside the CD that cannot radiate outside the CD.
\begin{figure}
\centering
\includegraphics[width=.7\textwidth]{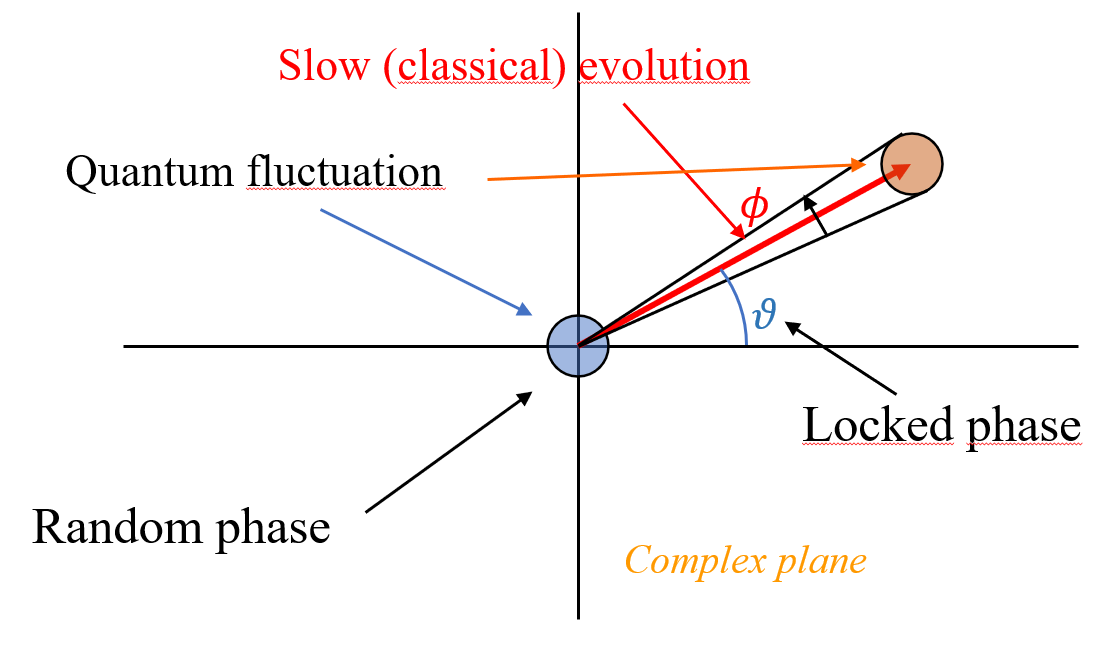}
\caption{Graphical representation of the quantum phase lock for a quantum coherent field. The disk at the origin represents the quantum fluctuation of a  field with zero expectation value whereas the disk on the right represents the quantum fluctuations for the field with a non-trivial order parameter $\phi$.  In the first case the quantum phase $\theta$ is completely indeterminate while in the second it is fixed by the temporal evolution of the order parameter represented by the red arrow that satisfies some classical field equations with an uncertainty $\Delta\theta=\frac1{\sqrt N}$}.
\label{fig:coherentWF}
\end{figure}
\item The symmetry breaking develops a negative energy gap which in the case of electrons and protons can be as large as 1 eV per particle. Such a large gap helps the coherent state to resist to decoherence due to thermal photons coming from the environment that are not able to transfer to the single particle enough energy to overcome the energy gap.
\item The quantum phase lock of the fields implies that the quantum amplitudes do not suffer from phase randomization, implying that cross sections and transition rates are proportional to the {\it square} of the number of particles (see Appendix \ref{app:incvscoh}).
\item The stability of the coherent states completely changes the kinematics of the reactions due to two main reasons:
\begin{itemize} 
\item a) the mass associated to the coherent state is the sum of the masses of its constituents
\item b) the presence of an energy gap that must be taken into account in the kinematics of the processes considered modifies the energy balance of the scattering (Fig. \ref{fig:coherentscatteringspectrum}).
\end{itemize}
Therefore in general new processes are possible thanks to a different kinematics. This also accounts for the substantial modification of the decay branching ratios when compared to the ``vacuum'' branching ratios.
\item The coherent interaction between coherent states can describe the transition to final states without the generation of highly energetic particles since the energy exchanged can be efficiently shared among a very large number of particles that can subsequently be dissipated as heat.

\end{itemize}

The theory can be considered as an extension of QED in the high-density limit and is therefore very general \cite{dickeham,prepDDfusion}. Is has been successfully applied 
to a number of condensed-matter physical problems like super-fluidity of $^4He$ and $^3He$ \cite{suphe4,he3}, crystal formation \cite{solidhe4}, Moessbauer scattering \cite{moessprep}, ferromagnetism, superconductivity, thermodynamics of liquid water\cite{watercoh}, cross-section amplification in gravitational antennas \cite{gravprep}, coherent neutrino elastic scattering \cite{prepneutrino}, biology \cite{emiliobiol}.
I will apply these ideas to the problem of production of neutrons via interactions between coherent states.

The quantum fields that will be considered are the $s-$wave electron field originated by the electrons previously belonging to the hydrogen atoms loaded into the metal, the proton field and the final-state fields: neutron field and neutrino field. The r\^ole of the lattice is in this context to host the electron and proton plasmas and will not be taken into account in this analysis. However, as described thoroughly in \cite{ref5}, the lattice has a fundamental role when it comes to the very existence of the electron and proton quantum plasmas and their oscillation properties (see in particular Chapters 5 and 8).

To this end, following the formalism of \cite{ref5,ref1,ref10} I define explicitly the many-particle quantum states of the various fields at play.
\subsection{Proton sector}\label{Proton sector}
I define the single-particle wave function of the absorbed protons as follows:
\begin{equation}
{\psi^p_{\vec q,\vec n,s}}\left( {\vec x,\vec \zeta } \right) =
\frac{ e^{i {\vec q \cdot \vec x}}}{\sqrt N_p}
\sum_{i=1}^{N_p}\phi^p(\vec x-\vec x_i)
\Braket{\vec\zeta | \vec n}
\chi_s
\label{eq:2.1}
\end{equation}
where 
$\vec x$ is the center-of-mass position of the scatterer, $\vec\zeta$ is the spatial elongation of the oscillator from the equilibrium position, $\vec q$ is the single-particle linear momentum, $\vec x_i$ is the equilibrium position of the $i^{th}$-scatterer,  $\phi^p(\vec x-\vec x_i)$ is the ground state 3d oscillator function of the protons
\begin{equation}
\phi^p(\vec x)=\frac{1}{\left(2\pi\right)^{3/4}r_{rms}^{3/2}}\exp\left(-\frac{\vec x^2}{4r_{rms}^2}\right)
\label{eq:2.2}
\end{equation}
where $r_{rms}=\frac1{\sqrt{2\omega_pm_p}}=0.086\angstrom\ll a_{x,y,z}$ is the width of the gaussian and $a_{x,y,z}=3.52\angstrom$ are the spacings of the crystal in the $x$, $y$, $z$ directions and $\Braket{\vec\zeta | \vec n}$ is the wave-function of the $\vec n^{th}$ excited state of the three-dimensional harmonic oscillator.

The wave function ${\psi^p_{\vec q,\vec n,s}}\left( {\vec x,\vec \zeta } \right) $ represents a single harmonic oscillator equally localized  in all the equilibrium sites of the lattice composed of $N$ sites.

The energy-momentum is given by
\begin{equation}
\begin{array}{l}
\Braket{\vec P}=
\int_{\vec x,\vec\zeta}
\psi^{p*}_{\vec q,\vec n,s}
\left( {\vec x,\vec \zeta } \right)
(-i\vec\nabla)
\psi^p_{\vec q,\vec n,s}
\left( {\vec x,\vec \zeta } \right)=
\vec q
\\
\Braket{E}=
\int_{\vec x,\vec\zeta}
\psi^{p*}_{\vec q,\vec n,s}
\left( {\vec x,\vec \zeta } \right)
H\left(\vec x,\vec \zeta\right)
\psi^p_{\vec q,\vec n,s}
\left( {\vec x,\vec \zeta } \right)=
e^2\frac {N_p}V \lambda_D^2+
\frac{\vec q^2}{2m_p}+(n_x+n_y+n_z)\omega_p,
\end{array}
\label{eq:2.4}
\end{equation}
where I have assumed the non-relativistic free single-particle energy spectrum.

The field operator of the proton plasma is given by
\begin{equation}
\Psi^p \left( {\vec x,\vec \zeta } \right) = 
\sum\limits_{\vec q,\vec n,s} 
\psi^p_{\vec q,\vec n,s}
\left( {\vec x,\vec \zeta } \right)
 a_{\vec q,\vec n,s} 
\label{eq:fieldop}
\end{equation}
with the anti-commutation relations
\begin{equation}
 \left\{a_{\vec q,\vec n,s},a^\dagger_{\vec q',\vec n',s'}\right\}=\delta_{\vec q \vec q'}\delta_{\vec n \vec n'} \delta_{ss'} 
\label{eq:2.3}
\end{equation}
and can be rewritten in terms of the wave-functions (\ref{eq:2.1}) as (I consider only the polarization along the $x$-axis)
\begin{equation}
\Psi^p \left( {\vec x,\vec \zeta } \right) = 
\sum\limits_{\vec q,s} 
{\left\{ 
	{{\psi ^p_{\vec q,\vec 0,s}}
		\left( {\vec x,\vec \zeta } \right)
		{a_{\vec q,\vec 0,s}} + 
		{\psi ^p_{\vec q,(1 0 0),s}}
		\left( {\vec x,\vec \zeta } \right)
		{a_{\vec q,(1 0 0),s}}	
} \right\}}
 \label{eq:wf}
\end{equation}

The operators $a^\dagger_{\vec q,\vec 0,s}$ and $a^\dagger_{\vec q,(1 0 0),s}$ create a fermion in the vibrating ground state and the first excited energy state. Only one polarization
has been chosen for simplicity.

Since I want to deal with a large number $N_p$ of protons I define the non-interacting state of the proton field in the following way.
Consider the two states ($k_F=\left(3\pi^2\frac {N_p}V\right)^{1/3}$ is the Fermi momentum)
\begin{equation}
\begin{aligned}
\Ket{0,\vec Q}_{pS}=
&\prod\limits_{|\vec q|\le k_F} 
\Ket{1,\uparrow,\vec q+\frac{\vec Q}{N_p},|\vec n|=0}\Ket{1,\downarrow,\vec q+\frac{\vec Q}{N_p},|\vec n|=0}
\cdot
\\
\cdot
&\prod\limits_{|\vec q|> k_F} 
\Ket{0,\uparrow,\vec q+\frac{\vec Q}{N_p},|\vec n|=0}\Ket{0,\downarrow,\vec q+\frac{\vec Q}{N_p},|\vec n|=0}
\end{aligned}
 \label{eq:2.5}
\end{equation}
and
\begin{equation}
\Ket{0,\vec Q}_{pP}=
\prod\limits_{\vec q} 
\Ket{0,\uparrow,\vec q,|\vec n|=1}\Ket{0,\downarrow,\vec q,|\vec n|=1}.
 \label{eq:2.6}
\end{equation}
I define the coherent state as
\begin{equation}
\Ket{\Omega_p,\vec Q}=
\prod\limits_{\vec q,s} 
\left[
\cos\theta
+
\sin\theta 
a^\dagger_{\beta_2}a_{\beta_1}
\right]
\Ket{\Omega,\vec Q}_{pert},
 \label{eq:grstate}
\end{equation}
where I have set $\beta_1=\{\vec q+\frac{\vec Q}{N_p},\vec 0,s\}$, $\beta_2=\{\vec q+\frac{\vec Q}{N_p},(1 0 0),s\}$ and $\Ket{\Omega,\vec Q}_{pert}=\Ket{0,\vec Q}_{pS}\Ket{0,\vec Q}_{pP}$. The state
$\Ket{\Omega,\vec Q}_{pert}$ is the \emph{unperturbed state}, composed
of the ensemble of protons in their vibrational ground state. 

A careful analysis carried out in \cite{prepfermioncoherence} shows that, due to the coherent interaction of the protons with the self-generated electromagnetic field, the state $\Ket{\Omega_p,\vec Q}$ (Eq. \ref{eq:grstate}) acquires an energy gap per particle of $\delta_p\simeq1$ eV with respect to the non-interacting state $\Ket{\Omega,\vec Q}_{pert}$ and that coherence is stable and conserved over domains (\emph{Coherence Domains}, CD) of diameter $D\sim10\mu m$ (corresponding to roughly 1 ng for Nickel), that fixes the number of coherent protons in a single CD to be
\begin{equation}
N_p\simeq\frac43\pi\left(\frac{D}{2a}\right)^3x=1.2\cdot 10^{13}x
\label{eq:protonnumber}
\end{equation}
where $x$ is the loading ratio of protons (I assume $x\sim1$ from now on).

It is important to note that the state  $\Ket{\Omega,\vec Q}_p$ carries a linear momentum $\vec Q$ that is equally shared among the $N_p$ protons. We have for the momentum
\begin{equation}
\Braket{\vec P}=
\int_{\vec x,\vec\zeta}
\quad
\Bra{\Omega_p,\vec Q}
\Psi^{p\dagger} 
\left( {\vec x,\vec \zeta } \right)
(-i\vec\nabla)
\Psi^p \left( {\vec x,\vec \zeta } \right) 
\Ket{\Omega_p,\vec Q}=\vec Q
\label{eq:2.9}
\end{equation}
and for the energy
\begin{equation}
\begin{aligned}
\Braket{E_{min}}=&
\int_{\vec x,\vec\zeta}
\quad
\Bra{\Omega_p,\vec Q}
\Psi^{p\dagger} 
\left( {\vec x,\vec \zeta } \right)
H\left(\vec x,\vec \zeta\right)
\Psi^p \left( {\vec x,\vec \zeta } \right) 
\Ket{\Omega_p,\vec Q}=
\\
=&N_p\frac{3(3\pi^2)^{2/3}}{10m_p}\left(\frac {N_p}V\right)^{2/3}+\frac{\vec Q^2}{2N_Pm_p}+
\frac 12N_pe^2\lambda_D^2\frac{N_p}V-N_p\delta_p
\end{aligned}
\label{eq:totalenergy}
\end{equation}

The first term in the RHS of Eq. (\ref{eq:totalenergy}) is the Pauli energy, the second term the kinetic energy, the third is the Coulomb energy and the fourth the energy gap due to the coherent interaction with the em field \cite{prepfermioncoherence}.
Note that the kinetic energy of the quantum state gets negligibly small in view of the large term $N_p$ in the denominator. The physical meaning is that the collection of protons behaves collectively as a single entity of mass $N_pm_p$. Due to the large value of $N_p$, kinetic energy may be neglected with respect to the other energetic contributions.

Let's make an estimate of the various contributions to the energy, taking the inter-atomic distance $a\simeq3.52\angstrom$. The density is
\begin{equation}
\frac {N_p}V=\frac {1}{a^3}=\left(\frac {1.98\cdot 10^{3}}{2.5}\right)^3 eV^{3}=1.78\cdot10^8 eV^{3}
\label{eq:density}
\end{equation}
and the Pauli energy is
\begin{equation}
E_{Pauli}=
N_p\frac{3(3\pi^2)^{2/3}}{10m_p}\left(\frac {N_p}V\right)^{2/3}=9.7\cdot10^{-4}N_p\quad eV
\label{eq:epauli}.
\end{equation}
As for the Coulomb interaction I take $\lambda_D\simeq \frac a2=1.76\angstrom=8.9\cdot 10^{-4} eV^{-1}$
\begin{equation}
\begin{aligned}
E_{Coulomb}=&
\frac12\alpha\int d^3\vec xd^3\vec y 
|\psi^p_{\vec q,\vec 0,s}(\vec x)|^2
|\psi^p_{\vec q,\vec 0,s}(\vec y)|^2
\frac{
e^{-\frac{|\vec x-\vec y|}{\lambda_D}}
}
{|\vec x-\vec y|}=
\\
=&
\frac12N_pe^2\lambda_D^2\frac{N_p}V
=N_p\frac{e^2}{8a}=6.3N_p\quad eV
\end{aligned}
\label{eq:ecoulomb}
\end{equation}
We conclude that, differently from electrons, the Pauli energy for protons is negligible compared to Coulomb's. The energy in Eq. (\ref{eq:totalenergy}) is therefore simplified to
\begin{equation}
\frac{\Braket{E_{min}}}{N_p}=
\frac 12e^2\lambda_D^2\frac{N_p}V-\delta_p
\label{eq:totalenergysimple}
\end{equation}
\subsection{Electron sector}
When the crystal is loaded with hydrogen additional electrons previously belonging to the hydrogen atoms are present in the lattice. I will call these electrons $s-$electrons.
Due to their smaller mass with respect to protons the $s-$electrons can be approximated as completely delocalized particles within the metal. One may be tempted to write for the single-particle wave-function
\begin{equation}
{\psi^e_{\vec k,\vec\vec n_e,s_e}}\left( {\vec x,\vec \zeta } \right)
 = \frac{e^{i {\vec k \cdot \vec x}}}{\sqrt{V}}
\Braket{\vec\zeta | \vec n_e}\chi_{s_e}
\label{eq:2.12old}\end{equation}
where 
$\vec x$ is the center-of-mass position of the scatterer, $\vec\zeta$ is the spatial elongation of the oscillator from the equilibrium position and $\vec k$ is the single-particle linear momentum.

However we must keep in mind that the protons present in the lattice are positively charged and influence the wave function of the electrons by attracting them. A proper treatment of such a problem would require the use of pseudopotentials and the solution of the mean-field associated problem. However I can make a few approximations that considerably simplify the formalism.

\begin{figure}
\centering
\includegraphics[width=.7\textwidth]{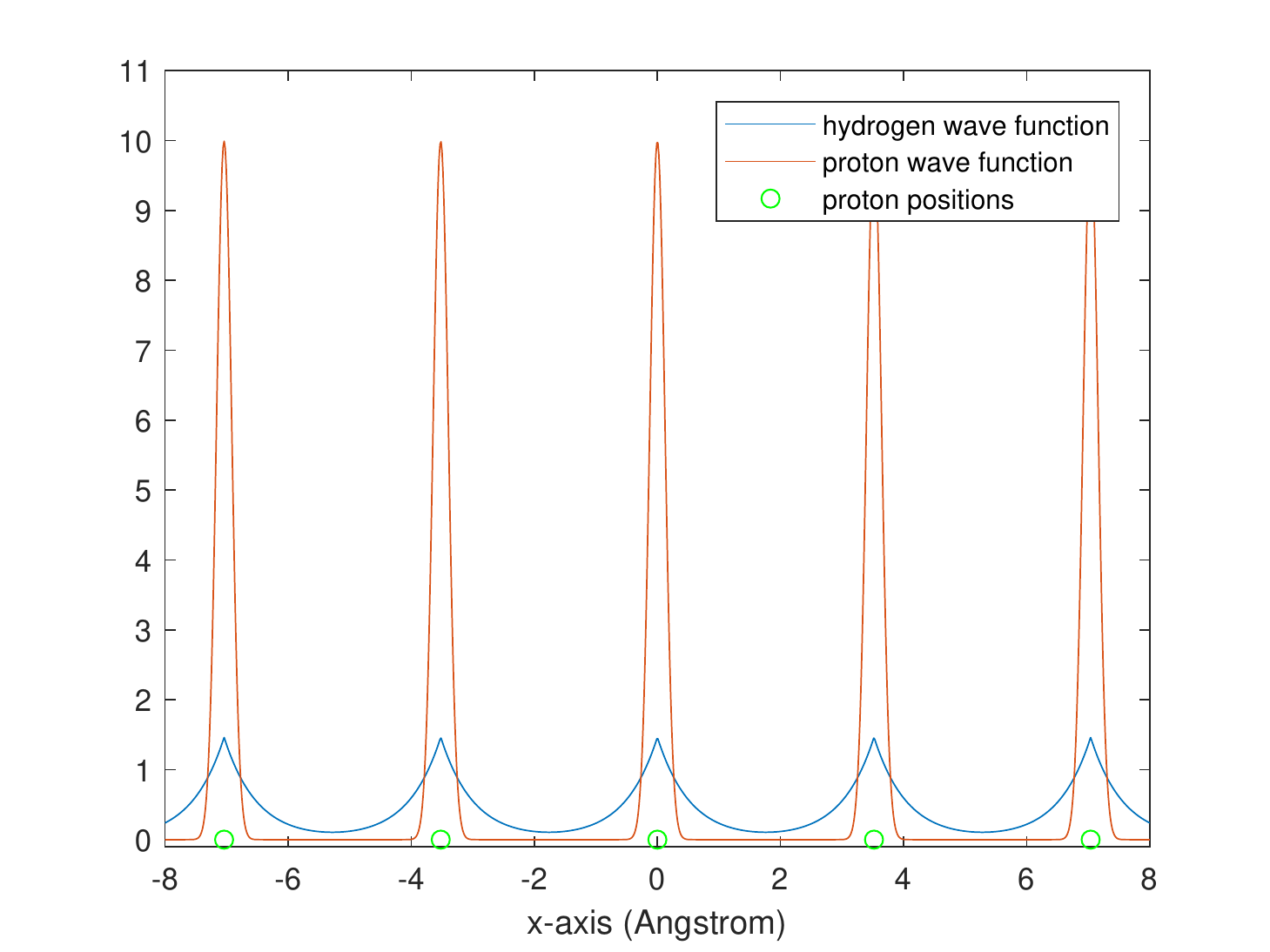}
\caption{Single-particle electron and proton wave functions}
\label{fig:electronWF}
\end{figure}

We start by saying that at very short distances between an electron and a proton the single-particle wave function must have a hydrogen-like character due to the strength of the Coulomb potential. We can approximate the wave-function to the ground state of the hydrogen atom, being that there is one $s-$electron available per proton ($N_e=N_p$).
The vibrational dynamics of the electrons will not be significantly affected, being the electrons much lighter than the protons and therefore able to accommodate their dynamics very quickly with respect to proton dynamics (Born-Oppenheimer approximation).
The resulting single-particle wave-function can be written as (Fig. \ref{fig:electronWF})
\begin{equation}
{\psi^e_{\vec k,\vec\vec n_e,s_e}}\left( {\vec x,\vec \zeta } \right)
 = e^{i {\vec k \cdot \vec x}}
\frac1{\sqrt{N_p}}
\sum_{i=1}^{N_p}
{\psi^{hydrogen}_{100}}\left( {\vec x-\vec x_i } \right)
\Braket{\vec\zeta | \vec n_e}\chi_{s_e}
\label{eq:2.12}
\end{equation}

Please note that, being the Bhor radius $a_0=5.3\cdot 10^{-11}$ m, the hydrogen wave-function has more than 99\% of the  integral of its square within the lattice elementary cell (whose side is $a\simeq 3.52\cdot 10^{-10}$ m ($4\pi\int_0^{3.52\cdot 10^{-10}}r^2dr\frac 1{\pi a_0^3}e^{-\frac{2r}{a_0}}\simeq 0.9998$). Such an approximation is therefore valid with good approximation.
The wave function ${\psi^e_{\vec k,\vec\vec n_e,s_e}}\left( {\vec x,\vec \zeta } \right)$ represents a single harmonic oscillator delocalized  in the whole lattice but spatially concentrated around the protons.

By repeating the arguments developed in section (\ref{Proton sector}) I can write for the electrons:
\begin{equation}
\frac{\Braket{E^{el}_{min}}}{N_e}=
\frac{3(3\pi^2)^{2/3}}{10m_e}\left(\frac {N_e}V\right)^{2/3}+
\frac 12e^2\lambda_D^2\frac{N_e}V-\delta_e
\label{eq:totalenergysimpleelectrons}
\end{equation}
where now the Pauli contribution cannot be neglected and amounts to $E_{Pauli}\simeq 1.8$ eV per particle.
\subsection{Excited proton states}
Once the existence of a coherent interacting state is established it is natural to ask ourselves what are the excitations of such a state. The answer is that there exist two types of excitations: incoherent and coherent.

Incoherent excitations are those for which single particles get removed from the coherent state and behave as single particles following a quasi-particle dispersion relation (Bogoliubov dispersion relation). In order for this to happen, it is necessary that some external excitation be sufficiently energetic to exchange enough energy to overcome the energy gap $\delta$ (Fig. \ref{fig:coherentscatteringspectrum}). This can happen for example via thermal excitation, where thermal photons are absorbed by the particle that is subsequently hurled into the thermal bath. Such an event has a probability given by the Boltzmann distribution and, given the large value of $\delta$, is negligibly small at temperatures in the range $0-1000$ C.

\begin{figure}
\centering
\includegraphics[width=.7\textwidth]{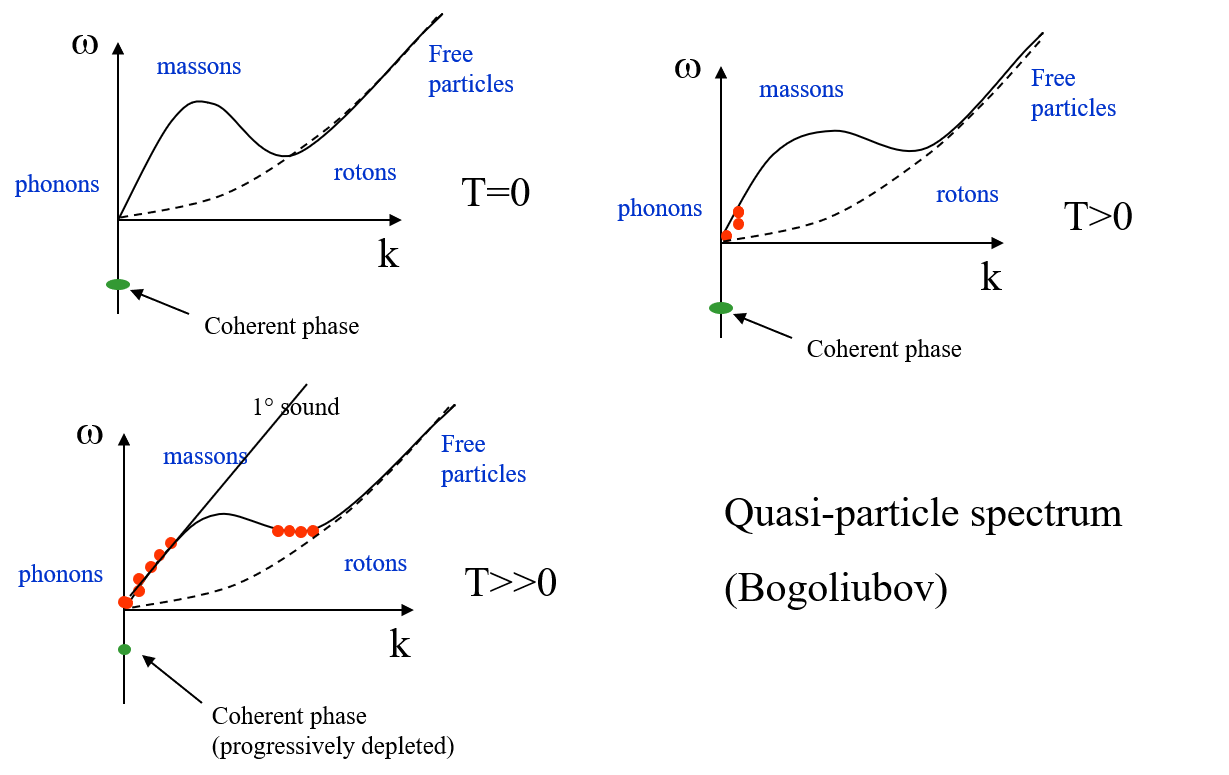}
\caption{Dispersion relation of the quasi-particles and of the coherent state at various temperatures. When temperature increases the coherent state gets progressively depleted in favor of the quasi-particle excitations, the energy gap decreases and eventually vanishes at a critical temperature (phase transition of the first or second kind, depending on the details of the process)}
\label{fig:coherentspectrum}
\end{figure}

When temperature is increased the number of incoherent excitations increases and a two-fluid picture emerges where there exist a coherent fraction $f_c$ composed of coherent protons and an incoherent (normal) fraction $f_n=1-f_c$ of non-coherent protons whose value is zero at $T=0$ and increases with increasing temperature (Fig. \ref{fig:coherentspectrum}).

Far more interesting are the coherent excitations, since they can give rise to \emph{coherent scattering}. Besides incoherent excitations we can think of a whole set of excitations of the matter sector that maintains the properties of coherence (energy lower than zero) but with an energetic content per particle higher than $E_{min}$.

In order to trigger these excitations it is necessary that some kind of pumping mechanism is present being able to keep the coherent excited states alive, that would be otherwise damped by various de-excitation mechanisms. Experimentally this can be accomplished by various means as electrochemical loading, heating, mechanical stress, laser irradiation to name a few. It is the task of the experimental activity to identify the most effective method to induce this type of excitations, possibly reaching a performance suitable for industrial applications.

For the sake of simplicity I will neglect for the moment thermal effects and restrict the analysis at zero temperature.
By looking at the structure of Eq. (\ref{eq:grstate}) I can write a coherent excited state as
\begin{equation}
\Ket{\Omega_p,\vec Q'}'=
\prod\limits_{\vec q,s} 
\left[
\cos\theta'
+
\sin\theta'
a^\dagger_{\beta'_2}a_{\beta'_1}
\right]
\Ket{\Omega,\vec Q'}_{pert},
 \label{eq:excstate}
\end{equation}
where $\vec Q'$ and $\theta'$ differ slightly from $\vec Q$ and $\theta$ to an amount that will be defined later on.
The state in Eq. (\ref{eq:excstate}) corresponds to an energy that is certainly higher than that on the state in Eq. (\ref{eq:grstate}) in that the mixing angle $\theta$ corresponds to the minimum of the energy.
\subsubsection{Initial and final states}
Since I am evaluating the transition probability for an electron capture in a metal, I have to consider that in the final state the number of protons and electrons is decreased by one unit.

In order to incorporate such a requirement I define the final state as:
\begin{equation}
\Ket{\Omega_{p, final},\vec Q}=
\frac 1{\sqrt{2N_p}}
\sum_{\substack{|\vec q'''|\le k_F,\\ \vec n'''s'''}}
a_{\beta_3}
\Ket{\Omega_p,\vec Q}'
 \label{eq:finalstate}
\end{equation}
where $|\vec q'''|\le q_F$ (in the following I will use the short-hand notation $\beta_3=\{\vec q'''+\frac{\vec Q}{N_p},\vec n''',s'''\}$). 
In Eq. (\ref{eq:finalstate}) I have applied a destruction operator on a single-particle state belonging to the state $\Ket{\Omega_{p},\vec Q}'$ and therefore contains $ N_p-1$ protons.  In the following, due to the very large value of $N_p$, I will make the approximation $N_p-1\simeq N_p$.

%
%

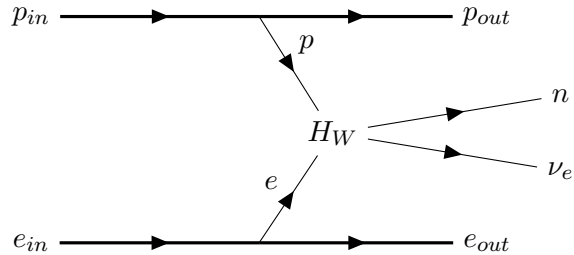
\begin{figure}
\centering
\begin{tikzpicture}
\begin{feynman}
\vertex (p1) {\(p_{in}\)}; 
\vertex[right=3cm of p1] (pv); 
\vertex[right=6cm of p1] (p2) {\(p_{out}\)}; 
\vertex[below=3cm of p1] (e1) {\(e_{in}\)}; 
\vertex[right=3cm of e1] (ev); 
\vertex[right=6cm of e1] (e2) {\(e_{out}\)};
\vertex[right=2cm of p1] (v1);
\vertex[right=2cm of e1] (v2);
\vertex[below right=1.25cm and 1.5cm of v1] (v3){\(H_W\)};
\vertex[above right=0.5cm and 3cm of v3] (n1) {\(n\)};
\vertex[below right=0.5cm and 3cm of v3] (nu1) {\(\nu_e\)};
\diagram* { {[edges=fermion]
(p1) -- [very thick] (pv) -- [very thick] (p2), 
(pv) --[edge label=\(p\)] (v3), (ev) -- [edge label=\(e\)] (v3),  
(e1) -- [very thick] (ev) -- [very thick] (e2),
(v3) -- (n1) , (v3) -- (nu1)},
};
\end{feynman} 

\end{tikzpicture}
\caption{Feynman diagram for the coherent neutron production}
\label{fig:cohEC}
\end{figure}

\subsection{Weak interaction matrix element}

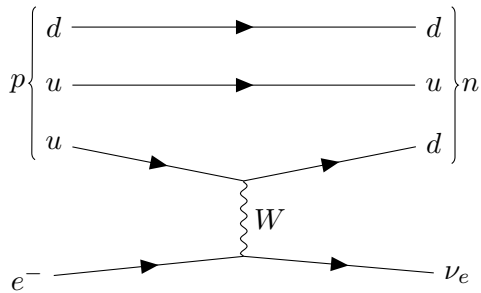
\begin{figure}
\centering
\begin{tikzpicture}
\begin{feynman}
\vertex (d1) {\(d\)}; 
\vertex[right=5cm of d1] (d2) {\(d\)}; 
\vertex[below=2em of d1] (u1) {\(u\)}; 
\vertex[right=5cm of u1] (u2) {\(u\)};
\vertex[below=2em of u1] (d3) {\(u\)}; 
\vertex[right=5cm of d3] (u3) {\(d\)};
\vertex[below right=0.5cm and 2.5cm of d3] (v1);
\vertex[below right=1cm and 0cm of v1] (v2);
\vertex[below right=0cm and 2.5cm of v2] (nu) {$\nu_e$};
\vertex[below left=0cm and 2.5cm of v2] (e) {$e^-$};
\diagram* { {[edges=fermion]
(d1) -- (d2),  (u1) -- (u2),
(d3) -- (v1) -- (u3), (e) -- (v2) -- (nu)},
(v1) -- [boson, edge label=\(W\)] (v2)
};
\draw [decoration={brace}, decorate] (d3.south west) -- (d1.north west) node [pos=0.5, left] {\(p\)};
\draw [decoration={brace}, decorate] (d2.north east) --  (u3.south east) node [pos=0.5, right] {\(n\)};
\end{feynman} 

\end{tikzpicture}
\caption{Feynman diagram of the quark spectator model for electron capture}
\label{fig:3}
\end{figure}

The matrix element is given by
\begin{equation}
\mathcal{T}=
\Bra{\Omega_{p, final},\vec Q_p}
\Bra{n,\vec q_n}
\Bra{\Omega_{e, final},\vec Q_e}
\Bra{\nu,\vec q_\nu}
\frac{G_F}{\sqrt2}
\int_{\vec x\vec\zeta}
j^\mu_h
\left( {\vec x,\vec \zeta} \right) 
j_{l\mu}
\left( {\vec x,\vec \zeta} \right)
\Ket{\Omega_e,\vec 0}
\Ket{\Omega_p,\vec 0}
 \label{eq:hadrlept}
\end{equation}
where $j^\mu_h \left( {\vec x,\vec \zeta} \right) $ and $ j_{l\mu} \left( {\vec x,\vec \zeta} \right)$ are the hadronic and leptonic currents, whose spinor component at the quark level is given by (Fig. \ref{fig:3})
\begin{equation}
\mathcal{M}=
\bar u_\nu \gamma^\mu(1-\gamma_5)u_e\bar u_d\gamma_\mu(1-\gamma_5)u_u.
 \label{eq:feynmanEC}
\end{equation}
The theoretical calculation of the EC decay rate, whose Feynman diagram is depicted in Fig. (\ref{fig:cohEC}) is a quite involved task. I will make use of a similar calculation that has been carried out in the case of muon capture by a proton in \cite{goaverts}.
Apart from the numerical values, the feature that must be incorporated in the result of reference \cite{goaverts} is the many-particle nature of the interaction.
The lepton capture decay rate  in \cite{goaverts} can be written in a simplified form as
\begin{equation}
\Gamma_{0,\text{lepton capture}}=
G_F^2V_{ud}^2 |\psi_s(\vec 0)|^2
\frac{(s-m_n^2)^2(3s+m_n^2)}{2\pi s^2}
 \label{eq:Gamma0}
\end{equation}
where $G_F$ is the weak Fermi constant, $\psi_s(\vec 0)$ is the single-particle wave-function of the electron at the proton position, $V_{ud}$ is the $u-d$ Cabibbo-Kobayashi-Maskawa matrix component, $s$ is the center-of-mass energy squared and $m_n$ the neutron mass.

We need now to find out how Eq. (\ref{eq:Gamma0}) gets modified by the multi-particle nature of the interaction. To do this I consider the sum of the contributions coming from all the states inside the Fermi surface. The momenta involved in the sum in Eq. (\ref{eq:finalstate}) are very small with respect to the masses of both the electrons and the protons allowing us to factor the spinor contribution out of the sums.

The hadronic component of the matrix element is given by
\begin{equation}
j^0_h
\left( {\vec x,\vec \zeta} \right) 
=
\Bra{\Omega_{p, final},\vec Q'}
\Bra{n,\vec q_n}
\Psi^{n\dagger} 
\left( {\vec x,\vec \zeta} \right) 
\Psi^p 
\left( {\vec x,\vec \zeta} \right) 
\Ket{\Omega_p,\vec Q}
 \label{eq:hadr}
\end{equation}

By substituting the explicit expressions the following amplitudes are relevant
\begin{equation}
\begin{aligned}
\Bra{0}&
a_{\beta_0} 
 a^\dagger_{\beta_3} 
a_{\beta} 
a^\dagger_{\beta_0} 
\Ket{0}=
\delta_{\beta_0\beta_3}\delta_{\beta\beta_0}
\\
\Bra{0}&
a_{\beta_0} 
 a^\dagger_{\beta_3} 
a_{\beta} 
a^\dagger_{\beta_2} 
 a_{\beta_1} 
a^\dagger_{\beta_0} 
\Ket{0}=
\delta_{\beta_0\beta_3}\delta_{\beta\beta_2}\delta_{\beta_1\beta_0}
\\
\Bra{0}&
a_{\beta_0} 
a^\dagger_{\beta'_1} 
 a_{\beta'_2} 
 a^\dagger_{\beta_3} 
a_{\beta} 
a^\dagger_{\beta_0} 
\Ket{0}=
\delta_{\beta_0\beta'_1}\delta_{\beta'_2\beta_3}\delta_{\beta\beta_0}
\\
\Bra{0}&
a_{\beta_0} 
a^\dagger_{\beta'_1} 
 a_{\beta'_2} 
 a^\dagger_{\beta_3} 
a_{\beta} 
a^\dagger_{\beta_2} 
 a_{\beta_1} 
a^\dagger_{\beta_0} 
\Ket{0}=
\delta_{\beta_0\beta'_1}\delta_{\beta'_2\beta_3}\delta_{\beta\beta_2}\delta_{\beta_1\beta_0}
\end{aligned}
\label{eq:hadr2}
\end{equation}
and the spatial component of Eq. (\ref{eq:hadr}) takes the form
\begin{equation}
j^0_h
\left( {\vec x,\vec \zeta} \right) 
=
\sqrt{2N_p}
\psi^{n*}_{\vec q_n,s}
\left( {\vec x,\vec \zeta } \right)
(c_p+s_p)
\left[c'_p
\psi^p_{\vec q_p+\frac{\vec Q_p}{N_p},\vec 0,s}
\left( {\vec x,\vec \zeta } \right)
+s'_p
\psi^p_{\vec q_p+\frac{\vec Q_p}{N_p},\vec 1,s}
\left( {\vec x,\vec \zeta } \right)
\right]
\label{eq:hadr2a}
\end{equation}
where I have defined $c_p=\cos\theta_p$, $s_p=\sin\theta_p$, $c'_p=\cos\theta'_p$, $s'_p=\sin\theta'_p$.
Analogously the lepton component takes the form
\begin{equation}
j^0_e
\left( {\vec x,\vec \zeta} \right) 
=
\sqrt{2N_e}
\psi^{\nu*}_{\vec q_\nu,s}
\left( {\vec x,\vec \zeta } \right)
(c_e+s_e)
\left[c'_e
\psi^e_{\vec q_e+\frac{\vec Q_e}{N_e},\vec 0,s}
\left( {\vec x,\vec \zeta } \right)
+s'_e
\psi^e_{\vec q_e+\frac{\vec Q_e}{N_e},\vec 1,s}
\left( {\vec x,\vec \zeta } \right)
\right]
\label{eq:lept2a}
\end{equation}
where all the variables have a meaning analogous to those for the proton case.

Once the spatial integrals are performed (see Appendix \ref{app:spatint}), Eq. (\ref{eq:hadrlept}) can be finally written as the sum of two terms ($N_e=N_p$):
\begin{equation}
\begin{aligned}
&
\frac {N_pG_FR
}{\sqrt {2\pi a_0^3}}
(c_p+s_p)
(c_e+s_e)
\left[
c'_pc'_eI_{00}+
s'_ps'_eI_{11}
\right]
\simeq
\\
\simeq
&\frac {N_pG_FR
}{\sqrt {2\pi a_0^3}}
I_{00}
(c_p+s_p)
(c_e+s_e)(c'_pc'_e+s'_ps'_e)
\end{aligned}
\label{eq:hadspat}
\end{equation}
The decay rate in Eq. (\ref{eq:Gamma0}) gets modified into (I set $\eta=R(c_p+s_p)
(c_e+s_e)(c'_pc'_e+s'_ps'_e)$)
\begin{equation}
\Gamma_{CD}=
\frac{(N_pG_Fm_nV_{ud})^2}{2\pi a_0^3}\eta^2
I^2_{00}f(x)
 \label{eq:Gamma1}
\end{equation}
where I have defined the invariant mass as $\sqrt s=m_p+m_e+\Delta E_{p,vib}+\Delta E_{e,vib}$, $x=\frac{\sqrt s}{m_n}$ and
\begin{equation}
f(x)=
\begin{cases}
\frac{(x^2-1)^2(3x^2+1)}{x^4}
 \quad & \text{for $x>1$}
\\
 0 \quad & \text{for $x\le 1$}
\end{cases}
\label{eq:f}
\end{equation}
that represents the production rate of neutrons in a single coherence domain. The center-of mass energy $\sqrt s$ is the energy difference between the final and initial quantum states of the proton and electron plasmas. The mass terms stem from the fact that the final states contain one proton and one electron less than the initial states.

Please note the quadratic dependence in $N_p$ of the interaction rate in Eq. (\ref{eq:Gamma1}) stemming from the coherent nature of the interaction (see also Appendix \ref{app:incvscoh}). Being $N_p\simeq 10^{13}$, the neutron production rate gets amplified by orders of magnitude, so that neutrons are readily produced as soon as the $s-$electron and proton coherent states get excited above energy threshold.

\section{Neutron production rate}
I am now in a position to compute the rate of neutron production in a single CD. 

Putting numbers into Eq. (\ref{eq:Gamma1}), I get  ($G_F=1.17\cdot 10^{-23}eV^{-2}$, $I_{00}=0.013$,  $N_p=1.2\cdot 10^{13}$, $m_n=939.57 \text{MeV}$, $\eta^2\simeq 0.1$, $a_0=2.68\cdot 10^{-4}\text{eV}^{-1}$) for the single CD,
\begin{equation}
\Gamma_{CD}=2\cdot 10^{3}f(\frac{\sqrt s}{m_n})\text{ eV}.
\label{eq:gammacdnum}
\end{equation}
From Eq. (\ref{eq:gammacdnum}) it is clear that the neutron production rate is controlled by the value of the function $f$ and depends on the magnitude of the excitation of the vibrational energy of the plasma coherent states of the protons and electrons. 

In order to have a neutron production it is necessary that $s>m_n^2$ or, put differently, that
\begin{equation}
\Delta E_{p,vib}+\Delta E_{e,vib}>m_n-m_p-m_e.
\label{eq:condition}
\end{equation}
Being $\Delta m=m_n-m_p-m_e=782$ KeV, the condition for neutron production is
\begin{equation}
\Delta E_{vib}=\omega_p(\sin\theta_p'-\sin\theta_p)+
\omega_e(\sin\theta_e'-\sin\theta_e)>
\frac{\Delta m}{N_p}\simeq 6.5\cdot10^{-8}\text{ eV}.
\label{eq:condition1}
\end{equation}
Eq. (\ref{eq:condition1}) shows that, at threshold, the variation of the angles $\theta_p$ and $\theta_e$ are of order $10^{-8}$, implying that the quantum configuration of the excited states differ very slightly from that of minimum energy.

Due to energy conservation, as soon as a neutron is produced, the excitation amplitude of the coherent plasmas gets reduced below threshold until a new excitation of the vibrating degrees of freedom is released. Calling $\braket{\Delta E_{vib}}$ the average vibration excitation energy and $P_{in, vib}$ the average power of coherent excitation of the vibrational degrees of freedom of the $s-$electrons and of the protons, energy conservation implies that
\begin{equation}
\frac{d}{dt}\braket{\Delta E_{vib}}=
-\Gamma_{CD}\Delta m+P_{in, vib}
\label{eq:balance1}
\end{equation}
and substituting Eq. (\ref{eq:gammacdnum}) into Eq. (\ref{eq:balance1}) (time is measured in eV$^{-1}$ units)
\begin{equation}
\frac{d}{dt}\braket{\Delta E_{vib}}=
-1.5\cdot10^{9}
f(\frac{\sqrt s}{m_n})
+P_{in, vib}
=
-
2.7\cdot10^{-8}
\left(
\braket{\Delta E_{vib}}-\Delta m
\right)^2
+P_{in, vib}
\label{eq:enbal}
\end{equation}
where I have expanded $f(x)=16(1-x)^2+O\left((x-1)^3\right)$ to second order around $x=1$. 
The transient solution of Eq. (\ref{eq:enbal}) is given by
\begin{equation}
x=1+\frac{1}{\alpha t+\frac1{x_0-1}}
\label{eq:enbala}
\end{equation}
where $\alpha=4\cdot 10^{16}$ Hz. 
Eq. \eqref{eq:enbala} implies that when the system is below threshold the initial condition is $x_0\sim0^+$ and the dynamics is slow. When the reaction has started we get $x_0>0$ and the very large value of $\alpha$ implies that the dynamics of the transient becomes extremely fast and the interaction is always at equilibrium.
This feature may explain the occasionally observed bursts of neutron emission in non-equilibrium experiments.

The stationary solution of Eq. (\ref{eq:enbal}) is given by
\begin{equation}
\braket{\Delta E_{vib}}-\Delta m
=6\cdot10^3
\sqrt{P_{in, vib}}.
\label{eq:enbal1}
\end{equation}
The kinetic energy of the produced neutron is given by the kinematics of the reaction. In the center-of-mass frame the momentum of the outgoing particles is given by
\begin{equation}
p_{n,\nu}^*=\frac{m_n^2}{2\sqrt s}\left(\frac s{m_n^2}-1\right)
\label{eq:pstar}
\end{equation}
and by substitution of the stationary solution in Eq. (\ref{eq:enbal1}) I get
\begin{equation}
p_{n,\nu}^*=6\cdot10^3\sqrt{P_{in, vib}}.
\label{eq:pstar1}
\end{equation}
To have an idea of the numbers let's imagine delivering 37 nW$=1.5\cdot 10^{-4}\text{eV}^2$ of coherent excitation power per CD (which in general will be only a fraction of the total power delivered to the CD): I get $p_n^*=0.9$ eV, meaning that the produced neutrons are essentially at rest with a kinetic energy of $4.4\cdot10^{-10}$ eV. In this case $x-1=7.9\cdot10^{-8}$ (Eq. (\ref{eq:enbal1})) and the production rate is $\frac{dN_n}{dt}=3\cdot 10^5$ neutrons per second per CD (Eq. (\ref{eq:gammacdnum})). These neutrons, being very slow, have an extremely large cross section of interaction with other nuclei and are completely absorbed by the nuclei of the hydride, in particular by the nuclei composing the lattice and the protons dispersed in it. 
However, in case of phenomena involving very fast transients of the lattice state (instantaneous power peaks), it may happen that the excitation of the coherent states is so large that the produced neutrons have a large momentum (see Eq. (\ref{eq:pstar1})) allowing for energetic neutron bursts escaping from the lattice.

Let's consider now a sample with mass $\sim 1$ gr. Considering that roughly speaking the atomic mass number of the nuclei of the lattice is $A\sim$50, the number of CDs is given by $N_{CDs}=\frac{N_{Av}}{AN_p}\simeq10^9$. The number of produced neutrons becomes $\frac{dN_n}{dt}=3\cdot 10^{14}$ neutrons/sec. Assuming that the produced neutrons interact mainly with the neighboring protons, the excess power associated is $P_{ex}=Q_{pn}\frac{dN_n}{dt}=106$ Watt where $Q_{pn}=2.2$ MeV is the binding energy of the deuteron and with an input power of coherent excitation of $P_{in}=37$ Watt. This type of process does not seem to be very efficient from an energetic point of view being that the theoretical maximum coefficient of performance (COP) is 2.2 MeV/ 0.782 MeV=2.8, assuming that {\it all} the delivered trigger power is converted into coherent excitation. In practical cases such a condition is hardly obtained. It seems therefore difficult to obtain useful amounts of energy by means of deuterium production. 

Much more promising for the energy sector is the reaction 
\begin{equation}
^3\text{He+n} \rightarrow^4\text{He}
\end{equation}
 where  $Q_{^3He^4He}$= 20.5  MeV which is more than 7 times more energetic than that for deuterium production. Such a reaction has been observed by D.Alexandrov \cite{alexandrov}, both as transmutation of $^3$He into $^4$He and production of excess energy.
A more accurate analysis should include the reactions of the neutrons with the nuclei of the lattice and eventually other dopant nuclei, giving rise to more energetic processes together with a certain amount of transmutations.

In conclusion, this calculation implies the production of various isotopes of the lattice nuclei and of a measurable amount of deuterium. 
I predict that, by inducing coherent excitations of the crystal lattice highly loaded with pure hydrogen after deloading of the crystal a certain amount of produced deuterium will be detected via mass spectrometry. No significant excess energy release is expected. Such an experiment would be a strong experimental confirmation of the presented theory and is currently underway in the LEDA/Brane laboratory.

\section{Discussion}

It is a common idea that nuclear reactions are not affected by the environment in which they occur because the energies involved are different by various orders of magnitude compared to thermal energies \cite{Gang_2008}.
However in some cases thermal \cite{210po,198au} or mechanical \cite{metzhag} effects associated with nuclear processes  have been observed showing that, at least in some circumstances, the nucleus is indeed affected by its surroundings. 
The theory presented is an attempt to describe a possible LENR process within the bounds of the Standard Model.

Even though  such an approach is able to account for the production of neutrons in matter, it is unable at this stage to justify the absence of energetic radiation that  is expected whenever a neutron is captured by a nucleus ($\gamma$ rays, X rays, energetic neutrons).
In order to address this issue it will be necessary to apply again the concept of quantum coherence to the by-products of the reaction. For example it is worth of note the possibility that also the produced neutron field can on its turn interact coherently with the various plasmas of the lattice thus enjoying different channels for the energy exchange that do not involve highly energetic particle emission. 

Another limitation of my analysis lies in the inability to identify precisely what is the mechanism for the excitation of the coherent states. I expect that there is more than one method of achieving this result, for instance via some electrochemical or electric means or by thermodynamic cycles and it is the duty of experimental art to find it.

In the present paper I analyze only one of the many possible coherent nuclear interactions in matter. For example in deuterated Pd the production of $^4$He is given by the interaction of deuterons present in the metal and most probably the mechanism involves the strong and not the weak force. A description of a possible coherent mechanism for $^4$He and heat production in a Pd crystal has been developed by G. Preparata in \cite{ref5}.

The analysis presented in this paper has been developed without considering the effect of temperature whose main effect is the reduction of the degree of coherence of the quantum plasmas. However, its effect is not expected to change dramatically the described scenario, being the thermal energy of the order of 0.025 $eV$ small compared to the energy gap of the order of 1 $eV$ that protects the coherent state from being destroyed by the thermal fluctuations. Once the contribution of thermal effects is taken into account, in addition to coherent interactions also incoherent interactions can take place accounting for the highly energetic radiation that has also been observed in several LENR experiments.

\section {Conclusions}
In this work I have shown a calculation of the production  of neutrons in a metal matrix highly loaded with protons via electron capture by the protons.  

The basic idea is that the vibrational coherent configurations of the electrons and protons in the metal have a collective energy which is sufficient to overcome the mass energy gap for one single EC. The result of such a calculation is that ultra-cold neutrons are indeed produced together with a low-energy neutrino flux. The rate of production of such neutrons is sufficient to produce detectable amounts of isotopes. An experiment aimed at detecting the production of deuterium nuclei starting from pure hydrogen interacting with Ti crystals is currently underway in our laboratory. 

The theory presented gives solid theoretical grounds for the development of a technology for the low-cost production of isotopes and nuclear waste remediation as well as presently non existing low-energy monochromatic electron neutrino sources.
 It is also imaginable developing in the future the nano-engineering of coherence domains in order to obtain very advanced genuine quantum reactions of which a (currently out of reach by human kind) wonderful example are living cells.

\section{Acknowledgments}
I am grateful to A.~Widom and Y.~Srivastava for illuminating discussions and for bringing my attention to the mechanism of electron capture as one possible cold nuclear process. I also thank A.~De Ninno for encouraging me in the publication of the present work.
Finally, I am grateful to G.~Modanese for reviewing the manuscript.

\appendix
\appendixpage

\section{Spatial integrals}\label{app:spatint}
The spatial integrals
\begin{equation}
\begin{aligned}
&
\int_{\vec x\vec\zeta}
\psi^{n*}_{\vec q_n}
\left( {\vec x+\vec \zeta } \right)
\psi^p_{\vec q_p+\frac{\vec Q_p}{N_p},\vec n_p}
\left( {\vec x,\vec \zeta } \right)
\psi^{\nu*}_{\vec q_\nu}
\left( {\vec x+\vec \zeta } \right)
\psi^e_{\vec q_e+\frac{\vec Q_e}{N_e},\vec n_e}
\left( {\vec x,\vec \zeta } \right)=
\\
&=
\frac{1}{V}
\frac{1}{N_p}
\sum_{i=1}^{N_p}
\sum_{j=1}^{N_p}
\int_{\vec x}
e^{i {(\vec q_p+\frac{\vec Q_p}{N_p}-\vec q_n+\vec q_e+\frac{\vec Q_e}{N_e}-\vec q_\nu) \cdot \vec x}}
{\psi^{hydrogen}_{100}}\left( {\vec x-\vec x_j } \right)
\cdot
\\
&\cdot
\phi^p(\vec x-\vec x_i)
\int_{\vec\zeta}
e^{-i (\vec q_n+\vec q_\nu)\cdot \vec\zeta}
\Braket{\vec\zeta | \vec n_p}
\Braket{\vec\zeta | \vec n_e}
\\
&=
\frac RV
\frac{1}{N_p}
\sum_{i=1}^{N_p}
e^{i {(\vec q_p+\frac{\vec Q_p}{N_p}-\vec q_n+\vec q_e+\frac{\vec Q_e}{N_e}-\vec q_\nu) \cdot \vec x_i}}
I_{\hat n_p\hat n_e}(\vec q_n+\vec q_\nu)
\sim
\\
&\sim
\frac RV
\delta_{\vec q_p+\frac{\vec Q_p}{N_p}+\vec q_e+\frac{\vec Q_e}{N_e},\vec q_n+\vec q_\nu}
I_{\hat n_p\hat n_e}(\vec q_n+\vec q_\nu)
\sim
\\
&\sim
\frac R{V^2}
(2\pi)^3\delta^3(\vec q_p+\frac{\vec Q_p}{N_p}+\vec q_e+\frac{\vec Q_e}{N_e}-\vec q_n-\vec q_\nu)
I_{\hat n_p\hat n_e}(\vec q_n+\vec q_\nu)
\\
&\sim
\frac R{V^2}
(2\pi)^3\delta^3(\vec q_p+\frac{\vec Q_p}{N_p}+\vec q_e+\frac{\vec Q_e}{N_e}-\vec q_n-\vec q_\nu)
I_{\hat n_p\hat n_e}(\vec q_n+\vec q_\nu)
\end{aligned}
 \label{eq:spat1}
\end{equation}
where I have defined
\begin{equation}
R=\int_{\vec x}{\psi^{hydrogen}_{100}}\left( {\vec x} \right)
\phi^p(\vec x).
\label{eq:R}
\end{equation}
Putting the numbers $r_{rms}^{p}=0.086\angstrom$, $a_0= 0.5\angstrom$ we have $R=0.31$.

\section{Harmonic oscillator eigenfunctions}
The first two eigenfunction for the harmonic oscillator are:
\begin{equation}
\begin{aligned}
\Braket{\vec\zeta | \vec 0}
=&
\left(
\frac{m\omega}\pi
\right)^{3/4}
e^{-\frac{m\omega \zeta^2}2}
\\
\Braket{\vec\zeta | \hat n}
=&
\left(
\frac{m\omega}\pi
\right)^{3/4}
\sqrt{2m\omega}
\vec\zeta\cdot\hat n
e^{-\frac{m\omega \zeta^2}2}
\end{aligned}
 \label{eq:harm1}
\end{equation}
The oscillator integral, being $\omega_p=0.28\sqrt{x}$ eV and $\omega_e=12$ eV and $m_p\omega_p=e\sqrt{\frac{m_pN_p}{V}}\gg m_e\omega_e=e\sqrt{\frac{m_eN_e}{V}}$, is calculated in the dipole approximation:
\begin{equation}
\begin{aligned}
I_{\vec 0\vec 0}=&
\int_{\vec\zeta}
e^{-i (\vec q_n+\vec q_\nu)\cdot \vec\zeta}
\Braket{\vec\zeta | \vec 0}_p
\Braket{\vec\zeta | \vec 0}_e
\simeq 
\left(
\frac{m_p\omega_p}\pi
\right)^{3/4}
\left(
\frac{m_e\omega_e}\pi
\right)^{3/4}
\int_{\vec\zeta}
e^{-\frac{m_p\omega_p \zeta^2}2}=
\\
=&
\frac{1}{\sqrt2\pi}
\left(
\frac{m_e\omega_e}{m_p\omega_p}
\right)^{3/4}
\\
I_{\vec 0\hat n}\simeq&
-i
\sqrt{2m_e\omega_e}
\left(
\frac{m_p\omega_p}\pi
\right)^{3/4}
\left(
\frac{m_e\omega_e}\pi
\right)^{3/4}
\int_{\vec\zeta}
[{ (\vec q_n+\vec q_\nu)\cdot \vec\zeta}]
[{ \hat n_e\cdot \vec\zeta}]
e^{-\frac{m_p\omega_p \zeta^2}2}=
\\
=&
-i\frac{3(m_e\omega_e)^{5/4}}{\pi(m_p\omega_p)^{7/4}}
(\vec q_n+\vec q_\nu)\cdot \hat n_e
\\
I_{\hat n\vec 0}\simeq&
-i
\sqrt{2m_p\omega_p}
\left(
\frac{m_p\omega_p}\pi
\right)^{3/4}
\left(
\frac{m_e\omega_e}\pi
\right)^{3/4}
\int_{\vec\zeta}
[{ (\vec q_n+\vec q_\nu)\cdot \vec\zeta}]
[{ \hat n_p\cdot \vec\zeta}]
e^{-\frac{m_p\omega_p \zeta^2}2}=
\\
=&
-i\frac{3(m_e\omega_e)^{3/4}}{\pi(m_p\omega_p)^{5/4}}
(\vec q_n+\vec q_\nu)\cdot \hat n_p
\\
I_{\hat n\hat n}\simeq&
-i
2\sqrt{m_p\omega_pm_e\omega_e}
\left(
\frac{m_p\omega_p}\pi
\right)^{3/4}
\left(
\frac{m_e\omega_e}\pi
\right)^{3/4}
\int_{\vec\zeta}
[{ \hat n_e\cdot \vec\zeta}]
[{ \hat n_p\cdot \vec\zeta}]
e^{-\frac{m_p\omega_p \zeta^2}2}
\simeq
\\
\simeq&
{ \hat n_e\cdot \hat n_p}
\frac{3\sqrt 2}{\pi}
\left[
\frac{
m_e\omega_e
}{m_p\omega_p}
\right]^{5/4}
=
6{ \hat n_e\cdot \hat n_p}
\sqrt{
\frac{
m_e\omega_e
}{m_p\omega_p}
}
I_{\vec 0\vec 0}\simeq
0.94
I_{\vec 0\vec 0}{ \hat n_e\cdot \hat n_p},
\end{aligned}
\label{eq:harmintegral}
\end{equation}
The terms $I_{ \vec 0\hat n}$ and $I_{\hat n\vec 0}$ are linear in $\vec q_n+\vec q_\nu$ and cancel after integration over $\vec q_\nu$ in the center-of-mass frame. We are left with $I_{\vec 0\vec 0}\simeq 0.013$ and $I_{\hat n_p\hat n_e}=0.012$ in the optimal case where $\hat n_e$ and $\hat n_p$ are parallel.

\section{Incoherent vs coherent amplitude}\label{app:incvscoh}
\begin{figure}
\centering
\includegraphics[width=.7\textwidth]{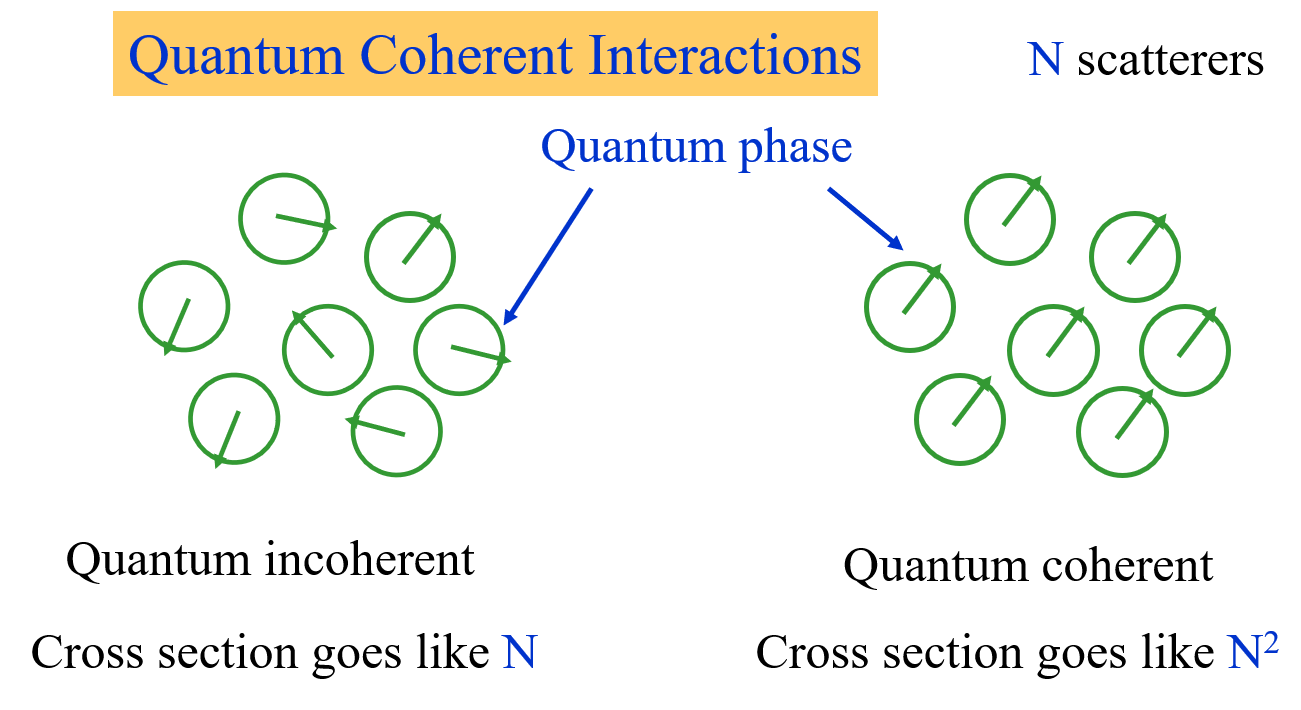}
\caption{Graphical schetch of incoherent and coherent summation of quantum phases}
\label{fig:phases}
\end{figure}
The $i-$th particle in the incoherent phase contributes to the total amplitude with a random phase $e^{i\theta_i}$. Calling $A_0$ the modulus of the amplitude associated to each particle I write the total amplitude as
\begin{equation}
A_{tot}=A_0\sum_i^Ne^{i\theta_i}.
\label{eq:inc1}
\end{equation}
The modulus of the amplitude is given by
\begin{equation}
|A_{tot}|^2=|A_0|^2
\int \frac{d\theta_1}{2\pi}
\int \frac{d\theta_2}{2\pi}
\dots
\int \frac{d\theta_N}{2\pi}
\left|\sum_i^Ne^{i\theta_i}\right|^2,
\label{eq:inc2}
\end{equation}
where I have averaged over each of the quantum phases of each particle owing to their incoherence. In order to evaluate the above integrals I use an iterative procedure by adding the contribution of one more particle. By defining
\begin{equation}
C^2_{N}=
\int \frac{d\theta_1}{2\pi}
\int \frac{d\theta_2}{2\pi}
\dots
\int \frac{d\theta_{N}}{2\pi}
\left|\sum_i^Ne^{i\theta_i}\right|^2,
\label{eq:inc3}
\end{equation}
I compute the contribution of the additional particle as
\begin{equation}
\begin{aligned}
C^2_{N+1}=&
\int \frac{d\theta_{N+1}}{2\pi}
\left|C_N+e^{i\theta_{N+1}}\right|^2=
\\
=&\int \frac{d\theta_{N+1}}{2\pi}
\left[C^2_N+1+2C_N\cos(\theta_{N+1})\right]=C_N^2+1.
\end{aligned}
\label{eq:inc4}
\end{equation}
Being $C_1=1$, I obtain $C^2_N=N$ and Eq. (\ref{eq:inc1}) can be written as
\begin{equation}
A_{tot}=\sqrt NA_0 e^{i\theta}
\label{eq:inc5}
\end{equation}
for some $\theta$. The rate of the process, which is proportional to the squared modulus of the amplitude, is therefore linear in the number of particles (see Fig. \ref{fig:phases}).

For example in an incoherent process the absorption rate of an incident beam of particles is proportional to the number of scatterers.

In the case of coherent amplitude all the phases are locked to a specific value evolving in time so that the total amplitude is given by
\begin{equation}
A_{tot}=A_0\sum_i^Ne^{i\theta(t)}=NA_0 e^{i\theta(t)},
\label{eq:inc6}
\end{equation}
implying that in coherent scattering the probability of interaction is enhanced by orders of magnitude with respect to the incoherent one due to the large number of particles involved in one single CD.

\bibliography{NeutronProductionViaElectronCaptureByCoherentProtonsArxiv1} 
\bibliographystyle{ieeetr}

\end{document}